\documentclass[twoside]{dis08}
\usepackage[latin1]{inputenc}
\usepackage[dvips]{graphicx,epsfig,color}
\usepackage{wrapfig,rotating}
\usepackage{amssymb,amsmath,array}

\usepackage{cite}

\newcommand{\gsim}{\raisebox{-0.07cm}{$\:\:\stackrel{>}{{\scriptstyle
 \sim}}\:\: $} }

\newcommand{\als}{\alpha_{\rm s}}
\newcommand{\as}{a_{\rm s}}

\def\cf{{C^{}_F}}
\def\nf{{n^{}_{\! f}}}

\begin{document}
\title{
{$\,$}\\[-6ex]
{\normalsize \sl  DESY 08-086, LTH 794 \hfill {$\,$}}\\[3ex]
NNLO Time-like Splitting Functions in QCD}

\author{S.~Moch$^1$ and A.~Vogt$^2$
%
%
\vspace{.3cm}\\
%
1-
Deutsches Elektronensynchrotron DESY \\
Platanenallee 6, D--15738 Zeuthen - Germany \\
\vspace{.1cm}\\
2-
Department of Mathematical Sciences, University of Liverpool \\
Liverpool L69 3BX - United Kingdom\\
}

\maketitle

\begin{abstract}
We review the status of the calculation of the time-like splitting functions 
for the evolution of fragmentation functions to the next-to-next-to-leading 
order in perturbative QCD.
By employing relations between space-like and time-like deep-inelastic 
processes, all quark-quark and the gluon-gluon time-like splitting functions 
have been obtained to three loops. 
The corresponding quantities for the quark-gluon and gluon-quark splitting at 
this order are presently still unknown except for their second Mellin moments.
\end{abstract}

\section{Introduction}
The transition from coloured quarks and gluons to colourless hadrons -- the 
so-called fragmentation or hadronization process -- is a Quantum Chromodynamics
(QCD) phenomenon with many important theoretical and phenomenological 
implications for the physics at high-energy colliders (see, e.g., Ref.~\cite
{trento:2008af} for a recent summary). 

\vspace{1mm}
For a given fractional momentum $x$ carried by the outgoing hadron $h$ the 
fragmentation functions (FFs) $\,D_{\! f}^{\,h}(x,Q^2)$ of the final-state 
partons $f$ obey evolution equations given by
\begin{equation}
\label{eq:Devol}
  \frac{d}{d \ln Q^2} \; D_{i}^{\,h} (x,Q^2) \:\: = \:\:
  \int_x^1 \frac{dz}{z} \; P^{\,T}_{ji} \left( z,\als (Q^2) \right)
  \:  D_{j}^{\,h} \Big( \frac{x}{z},\, Q^2 \Big) 
\end{equation}
where the summation over $\,j \, =\, q,\:\bar{q},\:g\,$ is understood. 
$Q^2$ is a time-like hard scale such as the squared four-momentum of the gauge 
boson in the reaction $\, e^+\,e^- \rightarrow\, \gamma\,,\,Z \,\rightarrow\, 
h + X$. 

\vspace{1mm}
The FFs are a basic ingredient at colliders, e.g., in the cross sections for 
particle production at high transverse momenta. 
However, being non-perturbative quantities, the $\,D_{\! f}^{\,h}$ have to be 
determined from global analyses (fits) of various experimental hadron and 
photon production data, if possible including flavour identification.
The available high-precision results for the FFs constrain and, eventually, 
improve the perturbative calculation of hadron and photon production in a way 
similar to global analyses of parton distribution functions (PDFs) for the 
initial protons. 
With a new kinematical range accessible in proton collisions at LHC energies
and at much increased luminosities, new opportunities for precision 
determinations of FFs will arise.
At the same time, improved evolution equations for the FFs beyond the 
next-to-leading order (NLO) of perturbative QCD will become important.

\section{Time-like splitting functions at three loops}
The scale dependence in Eq.~(\ref{eq:Devol}) is controlled by the time-like 
splitting functions $\,P^{\,T}_{ji}\,$ which admit an expansion in powers of 
the strong coupling $\als$, 
\begin{equation}
\label{eq:PTexp}
  P^{\,T}_{ji} \left( x,\als (Q^2) \right) \:\: = \:\:
  \as \, P_{ji}^{(0)\,T}(x) \: + \: \as^{\:\!2} \, P_{ji}^{(1)\,T}(x)
  \: +\: \as^{\:\!3} P_{ji}^{(2)\,T}(x) \: +\: \ldots \:\: ,
\end{equation}
where we normalize the expansion parameter as $\,\as \equiv \als(Q^2)/ (4\pi)$. 
The three-loop contributions $P_{ji}^{(2)\,T}$ are needed to complete
predictions to next-to-next-to-leading order (NNLO) accuracy in perturbative 
QCD.

\vspace{1mm}
The decomposition into flavour-singlet and non-singlet parts for the time-like
evolution equations (\ref{eq:Devol}) reads 
\begin{eqnarray}
\label{eq:Dnsevol}
  \frac{d}{d \ln Q^2}\:
  D^{\,\xi}_{\rm ns} \;
  &\!=\!&
  P_{\rm ns,\, \xi}^{\,T} \otimes
  D^{\,\xi}_{\rm ns}  
  \, ,
\\[1ex]
\label{eq:Dsgevol}
  \frac{d}{d \ln Q^2}
  \left( \begin{array}{c} \!D^{}_{\!S}\! \\ \!D_{\!g}\! \end{array} \right)
  \! &\!=\!& \! \left( \begin{array}{cc} 
         P_{\rm qq}^{\,T} & P_{\rm gq}^{\,T} \\[2mm] 
         P_{\rm qg}^{\,T} & P_{\rm gg}^{\,T}
   \end{array} \right) \otimes
  \left( \begin{array}{c} \!D^{}_{\!S}\! \\ \!D_{\!g}\! \end{array} \right)
  \quad \mbox{with} \quad
  D^{}_{\!S} \; \equiv \; \sum_{r=1}^{\nf} ( D_{q_r^{}} + D_{\bar{q}_r^{}} )
  \, ,
\end{eqnarray}
where $D^{\,\xi}_{\!ns}$ with $\xi = \pm,\rm v$ collectively denotes the
three types of non-singlet distributions $D_{ik}^{\,\pm} \, = \, D_{q_i} \pm 
D_{\bar{q}_i} \, - \, (D_{q_k} \pm D_{\bar{q}_k})\,$ and $\,D^{\,\rm v} 
\, = \, \sum_{r=1}^{\nf} (D_{q_r} - D_{\bar{q}_r})$. Here $\otimes$ abbreviates
the Mellin convolution written out in Eq.~(\ref{eq:Devol}), while $\nf$ stands 
for the number of effectively massless quark flavors. 

\vspace{1mm}
Motivated by the strong similarities between the evolution equations of the 
time-like FFs in Eq.~(\ref{eq:Devol}) and their space-like counter-parts for 
the PDFs, we have calculated in two publications~\cite{Mitov:2006ic,Moch:2007tx}
several of the splitting functions $P^{\,T}_{ji}$ entering the perturbative 
expansion in Eq.~(\ref{eq:PTexp}) to NNLO.
The leading-order (LO) terms in Eq.~(\ref{eq:PTexp}) are identical to the 
space-like case of the initial-state PDFs, a fact often 
referred to as the Gribov-Lipatov relation \cite{Gribov:1972ri}. 
Also the next-to-leading order contributions $\,P_{ji}^{(1)\,T}\!(x)$ have
been known for long~\cite{Curci:1980uw,Furmanski:1980cm,Floratos:1981hs} and
shown to be related to their space-like counter-parts by a suitable analytic 
continuation~\cite
{Curci:1980uw,Furmanski:1980cm,Floratos:1981hs,Stratmann:1996hn}.

\vspace{1mm}
Following these ideas, we have derived all three-loop non-singlet functions
corresponding to $D^{\,\xi}_{\!ns}$ and the diagonal singlet quantities 
$P_{\rm qq}^{\,(2),T}$ and $P_{\rm gg}^{\,(2),T}$ from the space-like NNLO 
results computed in Refs.~\cite {Moch:2004pa,Vogt:2004mw}.
To that end, we have applied two independent methods. 
One approach relied on mass factorization, knowledge of the infrared 
singularities and an analytic continuation in the momentum fraction from $x$ to 
$1/x$. The other approach implemented the concept of universal (kinematics 
independent) splitting functions of Refs.~\cite
{Dokshitzer:1995ev,Dokshitzer:2005bf}.

\vspace{1mm}
The first method started from the unrenormalized partonic structure functions 
in deep-inelastic scattering (DIS)~\cite{Moch:2004pa,Vogt:2004mw} in 
dimensional regularization. 
After iteratively subtracting virtual contributions due to the quark or gluon 
form factors~\cite{Moch:2005id,Moch:2005tm} (which are well known also in 
the time-like region) we applied an analytic continuation to the $x$-dependent 
real-emission functions $\,{\cal R}_{\,n}$ from $x$ to $1/x$ at each order in 
$\as$ similar to Refs.~\cite
{Curci:1980uw,Furmanski:1980cm,Floratos:1981hs,Stratmann:1996hn}.
During this step we have also taken into account the (complex) continuation of 
the DIS scale $q^2$ and the additional $D$-dimensional prefactor $x^{D-3}$ 
originating from the phase space of the detected parton in the time-like case 
(see, e.g., Refs.~\cite{Rijken:1996ns,Mitov:2006wy}). 
The subtle point in all this is the treatment of logarithmic singularities for 
$x \rightarrow 1$ starting with
\begin{equation}
\label{eq:l1xAC}
  \ln (1-x) \:\:\rightarrow\:\: \ln (1-x) - \ln x + i\,\pi \:\: .
\end{equation}
 
These steps lead to the (unrenormalized) one-parton inclusive fragmentation 
function in time-like kinematics, which can be re-assembled order by order in 
$\as$, keeping the real parts of the continued functions ${\cal R}_{\,n}$ only. 
Finally the time-like splitting functions (and coefficient functions) can be 
extracted iteratively from the mass factorization relations. 
Up to two loops all these the steps have been checked by a direct calculation 
for one-parton inclusive electron-positron annihilation~\cite{Mitov:2006wy}.

\vspace{1mm}
At three loops we have several consistency conditions, for instance the 
cancellation of all higher poles in dimensional regularization. 
Most importantly, sum rules exist for the first moment of 
$P^{\,(2),T}_{\rm ns, -}$ (number of fermions) and the second moment of 
$P_{\rm gg}^{\,(2),T}$ at $\nf = 0$ (momentum in pure gluodynamics).
These indicate different coefficients for the terms 
$C_{\!F}^{\,3}\,p_{\rm qq}(x)\,\pi^2\,\ln^2x\,$ of $\,P_{\rm qq}^{\,(2),T}$ 
and 
$C_{\!A}^{\,3}\,p_{\rm gg}(x)\,\pi^2\,\ln^2x\,$ of $\,P_{\rm gg}^{\,(2),T}$, 
where $p^{}_{ij}(x)$ are proportional to the LO splitting functions. Such 
differences are not unexpected in view of the imperfect real-virtual separation
at three loops in the approach described above.

\vspace{1mm}
As an independent confirmation of our analytic continuation approach including 
the sum-rule fix we thus needed a second, completely independent method. 
For this purpose we have adopted the relations between DIS and fragmentation 
processes of Ref.~\cite{Dokshitzer:1995ev} which exploit ideas of universal, 
i.e., kinematics independent parton-parton splittings based on a physical 
picture of strongly ordered life-times of successive parton fluctuations.
Introducing universal (non-singlet) splitting functions 
$P_{\,\rm ns}^{\,\rm univ}$, postulated to be identical for the time-like and 
space-like cases, Eq.~(\ref{eq:Devol}) can be represented in the following 
manner~\cite{Dokshitzer:2005bf}
\begin{equation}
\label{eq:DMSevol}
  \frac{d}{d \ln Q^2} \: D_{\rm ns}^{\sigma} (x,Q^2) \; = \;
  \int_x^1 \frac{dz}{z} \: 
  P_{\rm ns}^{\rm univ} \left( z,\als (Q^2) \right) \:  
  D_{\rm ns}^{\sigma} \Big( \frac{x}{z},z^{\sigma}Q^2 \Big) \:\: ,
\end{equation}
where the notation covers both the (time-like $q$, $\sigma=1$) FFs and the 
(space-like~$q$, $Q^2 \equiv - q^2$, $\sigma=-1$) PDFs. 
An additional shift in the argument of $\als$ in Eq.~(\ref{eq:DMSevol}) also
included in~\cite{Dokshitzer:2005bf} is irrelevant for our discussion.

\vspace{1mm}
The $\als$-expansion of Eq.~(\ref{eq:DMSevol}) correctly accounts
for the NLO difference 
$\delta\:\! P_{\rm ns}^{\,(1)}(x) = P^{\,(1),\, \sigma=+1}_{\rm ns}(x) - P^{\,(1),\,\sigma=-1}_{\rm ns}(x)$ 
between the space- and time-like splitting functions 
\cite{Curci:1980uw,Furmanski:1980cm,Floratos:1981hs}. 
At NNLO it leads to the remarkably compact expression 
\begin{equation}
\label{eq:dP2DMS}
  \delta\:\! P_{\rm ns,\, \xi}^{\,(2)}(x) \; = \;
   2 \left\{ \Big[ \ln x \cdot \widetilde{P}_{\rm ns,\, \xi}^{\,(1)} \Big] 
              \otimes P_{\rm ns}^{\,(0)}  
          + \Big[ \ln x \cdot P_{\rm ns}^{\,(0)} \Big]
              \otimes \widetilde{P}_{\rm ns,\, \xi}^{\,(1)} \right\}
\end{equation}
where $\,\xi = \pm,\: \rm v$, the Mellin convolution is again denoted by
$\otimes$, and 
\begin{equation} 
  2\,\widetilde{P}_{\rm ns,\, \xi}^{\,(n)}(x) \; = \; 
   P_{\rm ns,\, \xi}^{\,(n),\, \sigma=+1}(x) + P_{\rm ns,\, \xi}^{\,(n),\, \sigma=-1}(x) 
  \:\: .
\end{equation}
The evaluation of Eq.~(\ref{eq:dP2DMS}) exactly coincides with the result of 
the above analytic continuation including the correct coefficient for the term 
$\cf^{\! 3}\, p_{\rm qq}(x)\,\pi^2\,\ln^2x$ of $P_{\rm ns,\, \xi}^{\,(2),T\!}$,
as the required vanishing first moment is manifest in Eq.~(\ref{eq:dP2DMS}).
This result provides both the desired confirmation of the analytic continuation
and strong evidence in support of the ansatz~(\ref{eq:DMSevol}).

\vspace{1mm}
The physical picture leading to Eq.~(\ref{eq:DMSevol}) assuming strong ordering 
in the life-times of successive parton fluctuations can be applied in complete
analogy to the case of gluodynamics. 
Hence a relation similar to Eq.~(\ref{eq:dP2DMS}) can be derived for the
space- and time-like difference of the gluon-gluon splitting function $P_{\rm gg}$.
In particular, this reasoning again solves the above mentioned problem with 
the term $C_{\!A}^{\:\!3}\, p_{\rm gg}(x)\, \pi^2\, \ln^2x$ of 
$P_{\rm gg}^{\,(2),T}$, automatically  yielding the correct coefficient in 
agreement with the momentum sum rule for $\nf = 0$. 
Moreover both methods, the ansatz~(\ref{eq:DMSevol}) and the analytic
continuation, agree for all remaining terms proportional to $C_{\!A}^{\:\!3}$, 
$C_{\!A}^{\:\!2\,} \nf$ and $C_A n_{\!f}^2$ (i.e., those not involving gluon 
emission from quarks).

\vspace{1mm}
The analysis of the large- and small-$x$ limits of the time-like splitting 
functions displays a number of interesting features. 
The large-$x$ behavior of the diagonal quantities $P^{\,(2)T}_ {\rm qq}$ and 
$P^{\,(2)T}_ {\rm gg}$ is identical to that of their space-like counter-parts 
up to the sign of the sub-leading $\ln (1-x)$ contribution as predicted in
Ref.~\cite{Dokshitzer:2005bf}.
In contrast, the small-$x$ behaviour in the time-like region is markedly 
different as, e.g., $\,xP^{T}_ {\rm gg}$ receives double-logarithmic 
contributions (up to $\as^n \ln^{2n} x$) with large coefficients.
Despite large cancellations between leading and subleading terms, both singlet 
quantities $\,xP^{\,(2)T}_{\rm ps}\!$ and $\,xP^{\,(2)T}_{\rm gg}\!$ show a 
huge enhancement already at values of $x \gsim 10^{-3}$, see the plots in 
Ref.~\cite{Moch:2007tx}.
Although the FFs for final state hadrons generally do sample regions of larger 
$x$ than the corresponding PDFs for initial protons, it will be very 
interesting to investigate the range of perturbative stability in $x$ once the 
NNLO time-like splitting functions are known completely.

\section{Conclusions}
We have briefly summarized the current knowledge on the time-like splitting 
functions in massless perturbative QCD.
To NLO, the complete set of functions for the evolution kernel in
Eq.~(\ref{eq:Devol}) has been known for a long time
\cite{Curci:1980uw,Furmanski:1980cm,Floratos:1981hs}.
At NNLO accuracy, progress has been made by relating space-like and time-like 
deep-inelastic processes. Presently only the off-diagonal quantities 
$P_{\rm gq}^{\,(2),T}$ and $P_{\rm qg}^{\,(2),T}$ are still not completely 
known as a `naive' analytic continuation fails, unlike at two loops, for the 
$\pi^2$ contributions. 
Access to these functions hence requires an extension of the methods employed 
so far, either by extending the concept of universal splitting functions 
\cite{Dokshitzer:2005bf} beyond non-singlet quantities or by refining the 
technique based on mass factorization and analytic continuation~\cite
{Mitov:2006ic,Moch:2007tx}. 
In particular the latter approach also offers other applications, e.g., to 
Higgs decay mediated through top-quarks as in Ref.~\cite{Moch:2007tx} where 
highly non-trivial three-loop results of Ref.~\cite{Baikov:2006ch} were 
successfully confirmed in this manner.
Thus any progress on the method of analytic continuation beyond the 
state of the art presented here may have far reaching applications.


\begin{footnotesize}

\end{footnotesize}


\end{document}